\documentclass{nature}
\usepackage{graphicx}
\makeatletter
\let\saved@includegraphics\includegraphics
\AtBeginDocument{\let\includegraphics\saved@includegraphics}
\renewenvironment*{figure}{\@float{figure}}{\end@float}
\makeatother

\usepackage{color}
\usepackage{epstopdf}
\usepackage{bm}
\usepackage{dcolumn}
\usepackage{verbatim}
\usepackage{mathbbol}
\usepackage{amsmath}
\usepackage{placeins}
\usepackage{mathrsfs}
\usepackage{gensymb}
\usepackage{physics}
\usepackage{upgreek}
\usepackage{subcaption}
\usepackage{MnSymbol}
\usepackage{float}

\usepackage{lineno}
\usepackage[font=small, labelfont=bf]{caption}
\usepackage{siunitx} 

\title{Broadband ultrafast self-heterodyned chiro-optical spectroscopy}

\begin{document}

{\centering
\maketitle}

\begin{center}

Francesco Gucci$^{1\dagger}$, Andrea Iudica$^{1\dagger}$, Andres Valladares Y Tacchi$^{1\dagger}$, 
Andrea Schirato$^{1,2}$, Giulia Crotti$^{1}$, 
Ryeong Myeong Kim$^{3}$, Soo Min Lee$^{3}$, Jeong Hyun Han$^{3}$, 
Andrea Villa$^{1}$, Dawar Ali$^{4,5}$, Aurora Rizzo$^{4}$, 
Margherita Maiuri$^{1}$, Ki Tae Nam$^{3}$, 
Giuseppe Della Valle$^{1}$, Giulio Cerullo$^{1,6*}$.

\begin{affiliations}
\item Dipartimento di Fisica, Politecnico di Milano, Piazza Leonardo da Vinci 32, Milano 20133, Italy
\item Department of Physics and Astronomy, Rice University, 6100 Main St, Houston, Texas 77005, United States
\item Department of Materials Science and Engineering, Seoul National University, Seoul, Republic of Korea
\item CNR NANOTEC, Istituto di Nanotecnologia, c/o Campus Ecotekne, Via Monteroni, Lecce 73100, Italy
\item Dipartimento di Matematica e Fisica “Ennio De Giorgi”, Università del Salento, Via Monteroni, Lecce 73100, Italy
\item CNR-IFN, Piazza Leonardo da Vinci 32, Milano 20133, Italy
\item[]$^{\dagger}$ These authors contributed equally to this work.
\item[]$^{*}$ Corresponding author: giulio.cerullo@polimi.it
\end{affiliations}
\end{center}

\begin{abstract}
\textbf{Ultrafast chiro-optical spectroscopy provides unique access to the structural dynamics of molecules, spin-valley relaxation in semiconductors, and the non-equilibrium optical response of chiral nanophotonic systems and metasurfaces. Yet, because chiral signals are intrinsically weak and time-resolved spectroscopy probes small photoinduced changes, transient chiro-optical responses are often difficult to isolate from parasitic achiral contributions. Here, we introduce a broadband ultrafast chiro-optical spectroscopy technique that integrates a birefringent common-path interferometer with an optical polarization bridge to sensitively detect photoinduced changes in the polarization state of light. Phase-sensitive self-heterodyned detection enables simultaneous measurement of transient circular dichroism and optical rotatory dispersion across a broad spectral range with ultrafast temporal resolution. Balanced detection suppresses excess laser noise, enabling exceptional sensitivity ($<$50 {\micro}deg) which approaches the shot-noise limit. We demonstrate this approach on an ordered array of gold nano-helicoids, supported by a full-wave time-resolved model of the spatiotemporal dynamics of plasmonic non-equilibrium carriers and their associated optical nonlinearities. The model traces the system's transient chiro-optical response back to photoinduced modulations of the electric-magnetic dipole interaction in the nano-helicoid, elucidating the connection of near- and far-field dynamics in the non-equilibrium regime. We further investigate spin-selective carrier excitation, thermalization, and relaxation in a lead halide perovskite, establishing a novel approach to broadband time-resolved Faraday rotation. The simplicity, sensitivity, and wide applicability of this detection scheme provide a powerful platform for broadband ultrafast chiro-optical spectroscopy, opening new opportunities in biochemistry, solid-state physics, and nanophotonics.}
\end{abstract}


\section*{Introduction}\label{sec1}
Chirality is a pervasive feature of nature, manifesting itself in systems covering different scales, from the helical structure of DNA to the rotation of spiral galaxies. It refers to the intrinsic geometric property of an object that makes it non-superimposable on its mirror image. The two mirror image structures are called enantiomers, such as the left and right hands. Light is a powerful probe for detecting chiral signatures, as it can carry chiral properties itself. The simplest form of chiral light is a circularly polarized beam, where the two enantiomers are represented by the right ($\sigma^+$) and left ($\sigma^-$) handednesses\cite{ayuso_ultrafast_2022}. Circularly polarized light can also drive a chiral response in materials that, in the absence of photoexcitation, are achiral.

Over the past few decades, chiro-optical spectroscopy has emerged as a powerful tool for revealing structural information in biomolecules, enabling the enantioselective probing of molecular conformations\cite{changenet_recent_2023}. 
This early emphasis on biological systems was motivated by the central role of chirality in life sciences, where the homochirality of amino acids and sugars raised fundamental questions. 
However, the concept of chirality extends far beyond molecular systems, encompassing a diverse range of materials. 
Two-dimensional materials such as single-layer transition metal dichalcogenides represent a prototypical platform\cite{rong_interaction_2023, bao_light-induced_2022}, in which the interaction with circularly polarized light enables selective excitation of energy-degenerate valleys and generation of spin-polarized carriers, opening the field of valleytronics\cite{mak_control_2012, vitale_valleytronics:_2018, langer_lightwave_2018}. Furthermore, their use as building blocks for constructing van der Waals heterostructures, homobilayers and polaritonic systems has provided additional promising tuning knobs\cite{hubener_engineering_2021,tay_terahertz_2025,huang_giant_2023,zhu_creating_2024}. 

The advent of perovskites has further expanded the plethora of chiral light-matter interactions\cite{chen_ultrafast_2021}, spanning from the spin-selective excitation through circularly polarized light\cite{giovanni_highly_2015, giovanni_ultrafast_2019}, to the photogalvanic effect\cite{briscoe_photogalvanic_2025, ye_aboveroomtemperature_2014}, originating from Rashba-Dresselhaus splitting in the conduction band when the inversion symmetry is broken\cite{zhai_giant_2017, zheng_rashba_2015}. 

Nanophotonic systems, as a further illustrative example, can exhibit non-trivial chiro-optical responses arising either from their inherently chiral geometry\cite{fan_chiral_2012, kwon_chiral_2023, luo_plasmonic_2017, hentschel_chiral_2017}, or from extrinsic chirality induced by their spatial arrangement\cite{kim_enantioselective_2022}.
The latter mechanism is especially relevant in metasurface conurations, where refined control over orientation and placement of the nanostructures enables tailored chiral behaviour\cite{hentschel_chiral_2017, link_virtual_2021, wu_chiral_2022}. 
More recently, strong, topological fields carrying chirality have been shown to efficiently drive exotic phenomena, including e.g.~the Floquet-topological insulator state in graphene\cite{lesko_optical_2024}, the light-driven Haldane model in hBN\cite{mitra_light-wave-controlled_2024}, or the valley-selective bandgap engineering in centro-symmetric MoS$_2$\cite{tyulnev_valleytronics_2024}.
The control of chirality in condensed matter systems has also been extended to the phononic and structural degrees of freedom. Chiral and circular phonons and their light-induced driving offer a new way to break time reversal symmetry, thus generating pseudo-magnetizations through non-Maxwellian fields, in otherwise non-magnetic materials \cite{nova_effective_2017,basini_terahertz_2024}. 


The key observable in any chiro-optical spectroscopy technique is the complex chiro-optical susceptibility $\chi_\text{ch}(\omega)$, that can be expressed as $\chi_\text{ch}(\omega)\propto$ ORD($\omega$)+i CD($\omega$), with the optical rotatory dispersion (ORD) and the circular dichroism (CD) being connected via Kramers-Kronig relations. 
CD and ORD can be regarded, respectively, as the chiral counterparts of the absorptive and dispersive components of the complex achiral dielectric susceptibility $\chi_\text{ach}(\omega)$ of the system under study\cite{rhee_coherent_2012}. Thus, the CD (ORD) signal is the difference in absorption (refraction) between left (LCP) and right (RCP) circularly polarized light. The static CD is typically measured using a photoelastic modulator (PEM) to switch a monochromatic beam between LCP and RCP, and synchronously detecting the absorption changes using  a lock-in amplifier (LIA), achieving spectral resolution via a monochromator\cite{jasperson_improved_1969}. On the other hand, the static ORD can be accessed by measuring the rotation angle of the polarization plane of a linearly polarized beam, following its transmission through the sample.

Initially, time-resolved (TR) chiro-optical spectroscopy found its main application in the real-time tracking of the three-dimensional structure of (bio)molecules following photoexcitation\cite{lewis_new_1985, xie_picosecond_1989, changenet_recent_2023, hache_multiscale_2021, meyer-ilse_recent_2013, oppermann_broad-band_2019, oppermann_chiral_2022}. While this approach does not have the atomic scale structural sensitivity of other methods such as ultrafast X-ray/electron diffraction, still it combines experimental simplicity with the ability to work under physiological conditions in water solution. In materials science, TR-CD and TR-ORD (also known as time-resolved Faraday rotation) enable the study of the dynamics of spin-polarized carriers in hybrid organic/inorganic perovskites and of spin/valley scattering processes in two-dimensional transition metal dichalcogenides\cite{giovanni_highly_2015,giovanni_ultrafast_2019,bourelle_optical_2022,gucci_ultrafast_2024,dal_conte_ultrafast_2015}. The study of the ultrafast chiral response of nanophotonic systems and metasurfaces is also of interest\cite{crotti_giant_2024}. In addition, the ultrafast control of crystal chirality via nonlinear phononics \cite{zeng_photo-induced_2025} opens new directions for enantioselective steering of chemical reactions, information technology, and more broadly, the control and discovery of exotic phenomena in out-of-equilibrium quantum matter.

The measurement of the transient chiro-optical response of a sample is technically very demanding. Chiral signals, in fact, are already orders of magnitude smaller than the corresponding achiral ones, and since TR spectroscopy measures a weak photoinduced change, it gives rise to a very low signal, which is challenging to detect and distinguish from parasitic achiral photoinduced signals. The experimental approaches for TR chiro-optical spectroscopy can be classified into two groups: differential absorption and ellipsometry techniques. 
The former is based on a helicity-resolved pump-probe configuration, where the polarization of the probe light is switched on a pulse-to-pulse basis from LCP to RCP by means of a PEM, similar to the ones used in stationary CD. Over the years, considerable efforts have been devoted to improving versatility, performance and spectral coverage \cite{trifonov_broadband_2010, scholz_ultrafast_2019}, culminating in the setup by Oppermann et al.\cite{oppermann_ultrafast_2019}, which demonstrated exceptional sensitivity ($\sim$700 mdeg) in the UV range. However, this method relies on sequential measurements, making it more susceptible to laser fluctuations, and decreasing its signal-to-noise ratio. Other technical challenges arise from the need for broadband polarization modulators, and the non-trivial modulation schemes used to acquire the desired TR-CD. Additionally, it only measures the imaginary part of the transient chiro-optical susceptibility.

Ultrafast ellipsometry, on the other hand, aims at reconstructing the polarization state of linearly polarized light following interaction with a chiral sample. It is background-free and allows for measuring the TR-CD and/or ORD in a single shot\cite{eom_single-shot_2012}, making it intrinsically more sensitive and robust to intensity and phase fluctuations of the laser source. 
The method was initially demonstrated by Niezborala and Hache \cite{niezborala_measuring_2006}, who successfully recorded the TR-CD by introducing a variable phase retardation in the probe path. Nevertheless, the sequential acquisition reduces its sensitivity and the wavelength dependence of the phase retarder limits the setup to single-color operation.

Recently, Cho and coworkers \cite{rhee_femtosecond_2009, eom_broadband_2011} proposed an innovative approach for the measurement of the steady-state and transient chiro-optical response of a sample. This method starts from considering that, when a light field with a well-defined (e.g.~vertical) polarization impinges on a chiral sample, it generates a polarization in the material with both achiral (vertical) and chiral (horizontal) components. In turn, this polarization radiates a vertically polarized field, the achiral free induction decay (AFID), together with a much weaker horizontally polarized field, the chiral free induction decay (CFID). The CFID, selected by a polarizer, is then measured by heterodyne detection, recording its delay dependent interference with a local oscillator (LO) field. Cho's heterodyne scheme has proven successful in measuring both steady-state broadband CD/ORD\cite{eom_single-shot_2012}, and their TR counterparts\cite{hiramatsu_communication:_2015}. 
However, the need for an external LO makes this technique experimentally cumbersome and highly susceptible to phase fluctuations between the interfering beams.


Here we introduce a broadband ultrafast chiro-optical spectroscopy technique addressing the need of high-sensitivity simultaneous measurements of TR-CD and TR-ORD.
The setup expands on previously reported steady-state detection schemes\cite{preda_time-domain_2018, ghosh_broadband_2021}, now enabling chiro-optical detection with femtosecond temporal resolution.
The combination of a birefringent common-path interferometer (CPI) and a polarization bridge allows us to sensitively measure the photoinduced polarization rotation of light transmitted through the sample due to the CFID. Field-sensitive inteferometric detection allows simultaneous measurement of the TR-CD and TR-ORD spectra, enabling full reconstruction of the non-equilibrium chiro-optical response. Our self-heterodyned approach features broadband operation and high sensitivity, thanks to the balanced detection which cancels excess noise. This enables the measurement of sub-mdeg signals in unprecedented acquisition times, approaching the shot-noise limit and enabling the detection of signals as small as tens of microdegrees.

To showcase the potential of our technique, we first characterize the static and ultrafast chiro-optical response of a square lattice of chiral plasmonic nanoparticles. 
The experiments are supported by a time-resolved model that combines Jones calculus, used to express the system's ORD and CD, with numerical simulations of the space-time dynamics of plasmonic non-equilibrium carriers. This allows us to quantify the photoinduced nonlinear spectro-temporal changes in the nanostructure's chiro-optical response.  
We then investigate spin dynamics in a lead halide perovskite upon circularly polarized photo-excitation, and show that our approach enables us to disentangle the spin population thermalization and relaxation processes, thereby demonstrating a novel approach for broadband time-resolved Faraday rotation.

\section*{Results}
\subsection{Experimental setup}

The core of our broadband ultrafast chiro-optical spectroscopy technique is a simple, yet powerful self-heterodyned ellipsometric detection scheme. Simultaneous measurement of broadband CD/ORD signals is based on time-domain Fourier transform (FT) spectroscopy combined with balanced detection and heterodyne amplification using a birefringent CPI\cite{ghosh_broadband_2021}. 
This interferometer, known as Translating Wedge based Identical pulses eNcoding System (TWINS), uses a sequence of birefringent plates and wedges to enable control of the relative delay between two orthogonal polarizations with interferometric precision (down to $\approx 5$ as in the visible)\cite{brida_phase-locked_2012}.

\begin{figure}[h]
\centering
\includegraphics[width=1\textwidth]{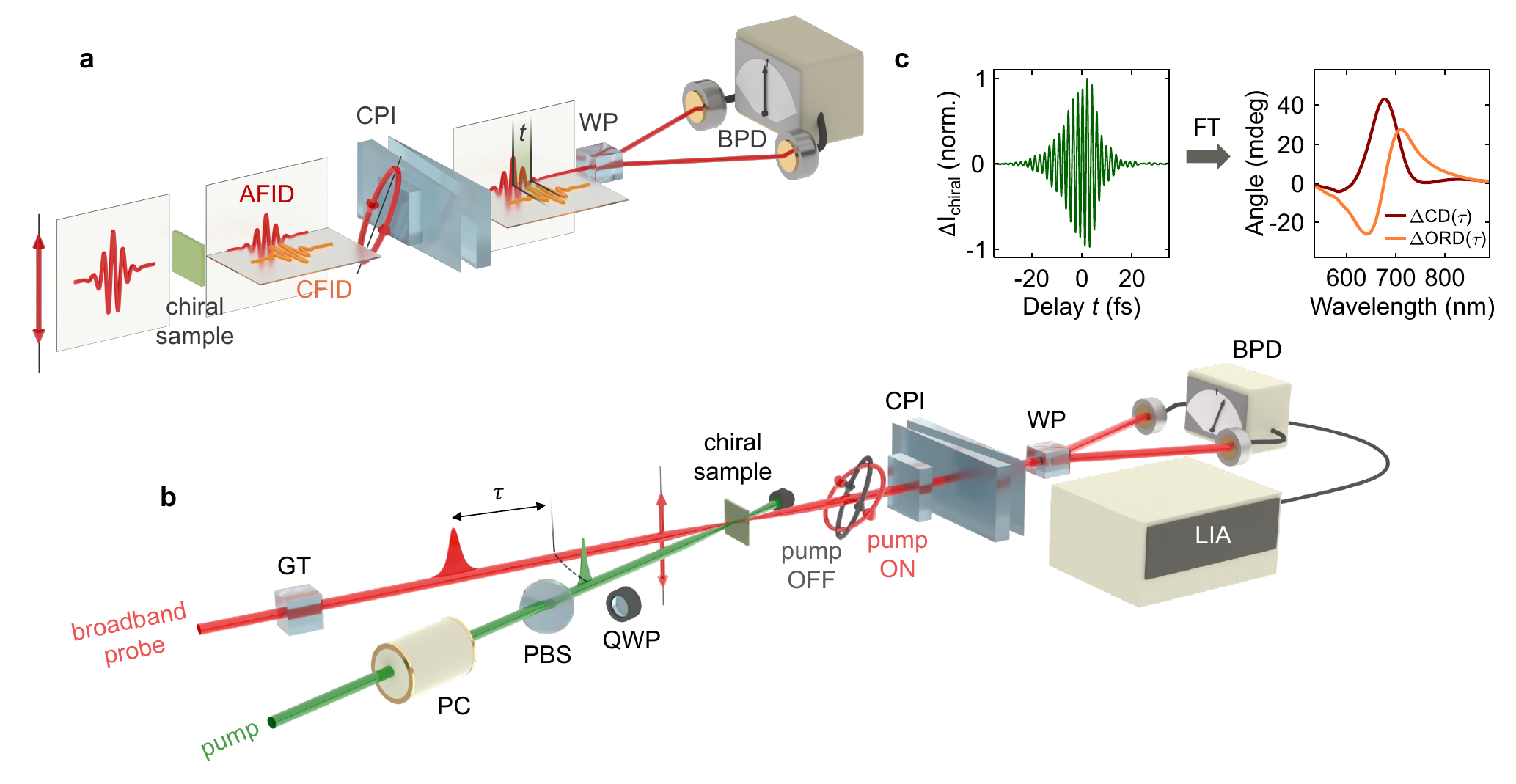}
\caption{\textbf{Ultrafast chiro-optical spectroscopy setup. a.} Working principle of the static chiro-optical measurements. After transmission through a chiral sample, the electric field of a linearly polarized probe pulse can be described as the sum of two components: achiral (AFID) and orthogonally polarized chiral (CFID) free induction decay, where the latter is much lower in amplitude (exaggerated in the figure for visualization). A common-path interferometer (CPI), constituted by a birefringent plate and a pair of birefringent wedges, scans the delay ($t$) between the two components. The combination of a Wollaston prism (WP) and a balanced photodetector (BPD) enables the detection of polarization rotation, measuring chiral interferograms upon insertion of the wedges. \textbf{b.} Ultrafast chiro-optical setup. The pump pulse repetition rate is halved by a Pockels cell (PC) and a polarizing beam splitter (PBS). A quarter wave plate can be inserted in the pump pulse path for circular polarization. A broadband probe pulse is linearly polarized by a Glan Taylor (GT) polarizer and delayed with respect to the pump by $\tau$. The detection scheme follows the procedure described in \textbf{a}, except that here we use a lock-in amplifier (LIA) to demodulate the signal. This allows us to extract the pump-induced change in the chiral interferogram, which we refer to as the differential chiral interferogram, $\Delta$I$_{chiral}$($t$,$\tau$). \textbf{c.} Example of a differential chiral interferogram and the associated $\Delta$CD and $\Delta$ORD spectra.}\label{fig1}
\end{figure}

In our approach, schematically sketched in Fig.\ref{fig1}a, the sample is illuminated by a broadband vertically polarized probe light which, upon interaction with the material, creates a polarization with both achiral (vertical) and chiral (horizontal) components. In turn, this polarization radiates a vertically polarized AFID, together with a much weaker horizontally polarized CFID, which rotates the polarization of the transmitted light field. This polarization rotation is sensitively detected by a so-called optical polarization bridge, which consists of a Wollaston prism (WP) splitting the light into two orthogonal polarization components of equal intensity, which are measured by a balanced photodetector (BPD), consisting of two photodiodes followed by a differential amplifier. The optical bridge measures any polarization rotation as an imbalance of the differential amplifier output. Frequency resolution is achieved by a FT approach, using the TWINS CPI to introduce an interferometrically stable delay $t$ between the horizontal CFID field and the vertical AFID, and recording a time-domain chiral interferogram,  I$_{chiral}$($t$). Its FT, after normalization by the FT of the achiral interferogram, gives the broadband stationary CD and ORD spectra (see Supplementary Note 1). Besides its simplicity and broadband operation, this approach achieves close to shot-noise-limited sensitivity, as it allows for cancellation of the common-mode light fluctuations of the probe with the balanced detector.

Here we extend this scheme to ultrafast detection of broadband differential CD and ORD ($\Delta$CD, $\Delta$ORD) by adding an ultrashort pump pulse and measuring, with a LIA, the pump-induced changes of the chiral interferogram of the probe. The experimental setup, schematically shown in Fig.~\ref{fig1}b, starts with an amplified ytterbium laser generating 200-fs pulses at 1030 nm and 100 kHz repetition rate. A fraction of the laser output is frequency doubled to generate a 515-nm pump pulse with arbitrary polarization (in our case, either linear or LCP/RCP), while another fraction is focused in a YAG plate to generate a broadband (550-950 nm) linearly polarized supercontinuum probe pulse, synchronized with the pump pulse. The pump is electro-optically modulated at half the repetition rate of the probe pulses via a Pockels cell (PC) coupled to a polarizing beam splitter. The probe polarization is controlled via a Glan-Taylor (GT) birefringent polarizer, having a contrast of 1:100000, to ensure a purely linear polarization state. The transmission axis of the GT polarizer is then carefully adjusted to align it with the extraordinary axes of the two wedges of the CPI. Pump and probe pulses are non-collinearly focused on the sample and the broadband probe is collected and sent to the polarization bridge (consisting of WP and BPD). A LIA, demodulating at the PC repetition rate, enables the detection of the pump-induced changes in the probe chiral interferogram 
as a function of the interferometrically stable delay $t$ between CFID and AFID, resulting in the differential chiral interferogram  $\Delta$I$_{chiral}$($t$, $\tau$), where  $\tau$ is the pump-probe delay.  As in the static case, we achieve frequency resolution by performing a FT with respect to $t$ of the differential interferograms (see Fig.~\ref{fig1}c, left panel) and obtain TR-CD and TR-ORD spectra for a given delay  $\tau$ (as shown in Fig.~\ref{fig1}c, right panel). The setup is able to record millidegree-level signals in few tens of seconds and its sensitivity can be further increased to $<$50 $\mu$deg (see Supplementary Note 2).

\subsection{Chiral plasmonic nanoparticles}

\begin{figure}[H]
\centering
\includegraphics[width=1\textwidth]{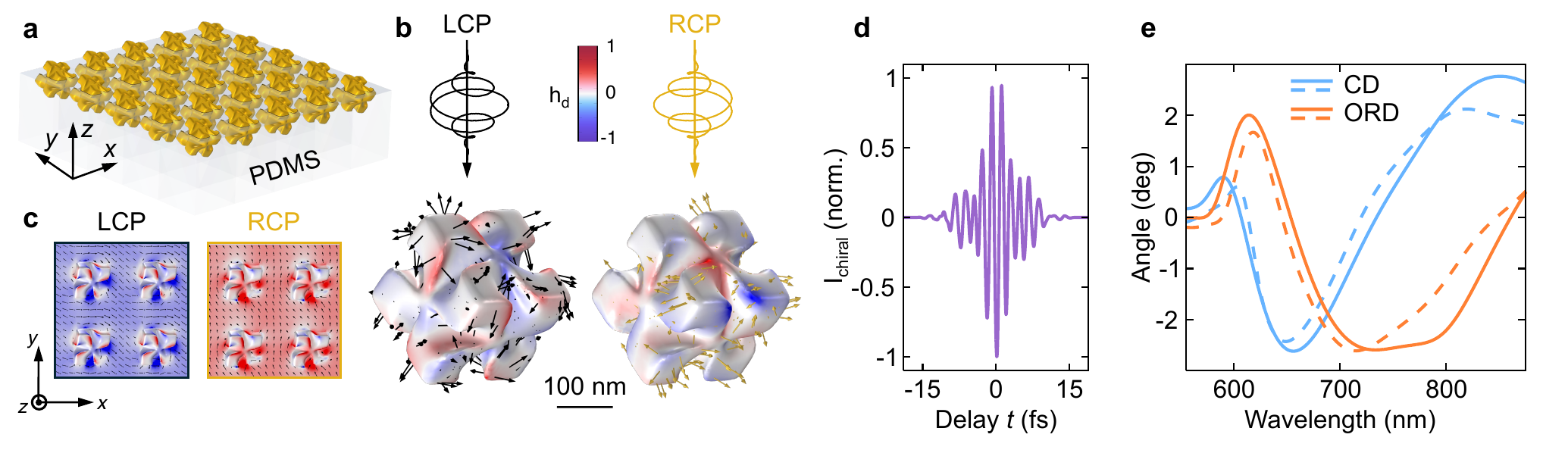}
\caption{\textbf{Static chiro-optical response of plasmonic helicoids. a.} Schematic of the system under study, consisting of an ordered square lattice of chiral Au nano-helicoids lying on a PDMS substrate. \textbf{b.} Spatial distribution of the (normalised) optical helicity density $h_d$ computed across one helicoid, evaluated at a wavelength of 655 nm in unperturbed conditions, for either LCP (left panel, black) or RCP (right panel, yellow) incoming light. Arrows indicate the electric field vectors, evaluated on the surface of the helicoid. \textbf{c.} Same as (\textbf{b}), shown as a top view (in a $xy$ plane at mid-height of the helicoids) across a few unit cells of the nanoparticle lattice. \textbf{d.} Experimental steady-state chiral interferogram. \textbf{e.} Measured (solid lines) and simulated (dashed lines) CD (blue) and ORD (orange) spectra of the plasmonic helicoid array in unperturbed conditions.}\label{fig2}
\end{figure}

We test our spectroscopy technique by performing both static and ultrafast chiro-optical measurements on a two-dimensional array of 432-symmetric chiral Au nano-helicoids.  
These nanoparticles are synthesized via a bottom-up approach, with handedness controlled by cetyltrimethylammonium bromide (CTAB) chirality transfer (see Methods Section 1) and were proven to exhibit a pronounced static chiro-optical activity\cite{lee_amino-acid-_2018}. 
The nanostructures are arranged in a square lattice of 400-nm periodicity on a polydimethylsiloxane (PDMS) substrate, as schematically depicted in Fig.~\ref{fig2}a.

We experimentally characterize the stationary chiro-optical response of the sample by Fourier transforming the steady-state chiral interferogram (Fig.~\ref{fig2}d), obtaining the CD and ORD spectra shown in Fig.~\ref{fig2}e as solid lines. 
The CD (blue solid trace) exhibits a resonant spectral profile, with a first positive peak near \SI{600}{nm}, a dip, which reaches its minimum at $\sim$\SI{650}{nm}, and a second peak, broader and higher in amplitude (exceeding 2 degrees), centered around \SI{850}{nm}.
The ORD (orange solid curve), in accordance with the Kramers-Kronig relations linking it to the CD, shows a positive band peaking around \SI{615}{nm}, followed by a broad negative dip in the \SI{700}{nm} -- \SI{800}{nm} range, and changes sign again beyond \SI{860}{nm}.
The results of our measurement are in good agreement with the CD spectrum recorded with a commercial spectrometer (Model Jasco J-1000, see Supplementary Note 3). 
We highlight that the chiral interferograms like the one shown in Fig.~\ref{fig2}d, besides giving access to both CD and ORD spectra, are recorded in remarkably short acquisition  times of $\sim 6$ seconds, namely more than one order of magnitude faster with respect to a commercial CD spectrometer ($\sim 2$ minutes). 

The pronounced chiral response of the nano-helicoids under study originates from their characteristic twisted geometry, which induces differential absorption of circularly polarized light depending on its handedness and wavelength. 
In particular, RCP light is absorbed more than LCP between \SI{605}{nm} and $\sim \SI{730}{nm}$ (resulting in negative CD), while the reverse occurs at shorter and longer wavelengths, where the CD becomes positive.
More fundamentally, the origin of CD in chiral systems is commonly attributed to interference between electric-magnetic dipole moments and electric dipole-quadrupole moments \cite{barron_molecular_2004}. 
In the case of our helicoids, a multipolar decomposition analysis reveals that a dominant contribution to the CD around 650 nm arises from the magnetic dipole moment, which results from the current loops induced on the curved surface of the nanoparticle (see Supplementary Note 4).
In essence, regions of opposite charge are formed across the chiral gaps, generating a capacitive coupling, similarly to that observed in split-ring resonators \cite{ahn_highly_2024}. Instead, the positive peak at longer wavelength originates from an electric dipole-quadrupole interaction \cite{kim_controlling_2021}.
 
The experimental characterisation of the sample's static chiro-optical behaviour is supported by a numerical model combining Jones calculus with full-wave simulations based on the finite element method.
This framework enables us to compute near- and far-field quantities, including the spatial distribution of the induced electromagnetic fields and the handedness-selective absorption.
Figure \ref{fig2}a presents a sketch of the simulated structure, consisting of an infinite square array of identical Au nano-helicoids, periodically arranged on a PDMS substrate. 
Since the orientation of individual nanoparticles in the experimental sample may vary, we assume in the model an effective unit cell configuration that best matches the measured spectra. 
Further details on the simulated geometry are provided in Methods Section 4.

Depending on the handedness of the incoming light (either LCP or RCP), the model predicts markedly different electromagnetic responses.
To quantify the handedness-selective light-matter interactions in the near field, we examine the optical helicity density $h_d = \text{Im} \big( \textbf{E} \cdot \textbf{H}^* \big)/ 2\omega c$, which quantifies the extent to which electric ($\textbf{E}$) and magnetic ($\textbf{H}$) fields locally wrap around the direction of light propagation at each point in space\cite{tang_enhanced_2011}. 
Figures \ref{fig2}b and \ref{fig2}c show the spatial distribution of the normalised $h_d$ at \SI{655}{nm}, evaluated on the nanoparticle surface and on the plane of the substrate (a $xy$ plane at mid-height of the helicoid), respectively, along with the three-dimensional orientation of the electric vector field (represented by the arrows).
In both cases, switching from LCP (black) to RCP (yellow) results in a clear sign inversion of $h_d$ at the nanoparticle's chiral hot spots and across the whole array, along with a reversal in the handedness of the induced electric field (arrows spin in the opposite direction) and an enhancement of the local twisting near regions of strongest chirality.

The simulations also provide access to the spatial distribution of the electric charge density along the helicoid surface (see Supplementary Note 4), further confirming the aforementioned strong magnetic dipole contribution underlying the structure's CD. 
To quantify the CD and ORD of the helicoid array, we solve Maxwell's equations to obtain the left and right co-polarized field transmission coefficients.
These are combined to retrieve the chiral susceptibility via intrinsically Kramers-Kronig-related expressions for the CD and ORD, which we derived from a Jones matrix formalism approach. 
Further details can be found in Methods Section 1 and in Supplementary Note 5.
The resulting CD and ORD are reported in Fig.~\ref{fig2}e as dashed lines, and compared with the measured spectra (solid lines). 
Considering the structural complexity of the system (at both the nano- and meso-scale), the simulations are in very good agreement with experiments, with only minor deviations (likely arising from the assumption of a perfectly periodic array of identical nanoparticles).
Overall, the numerical model accurately reproduces the key features observed experimentally in unperturbed conditions, and supports the mechanisms driving the system's chiral response.

\begin{figure}[H]
\centering
\includegraphics[width=0.7052\textwidth]{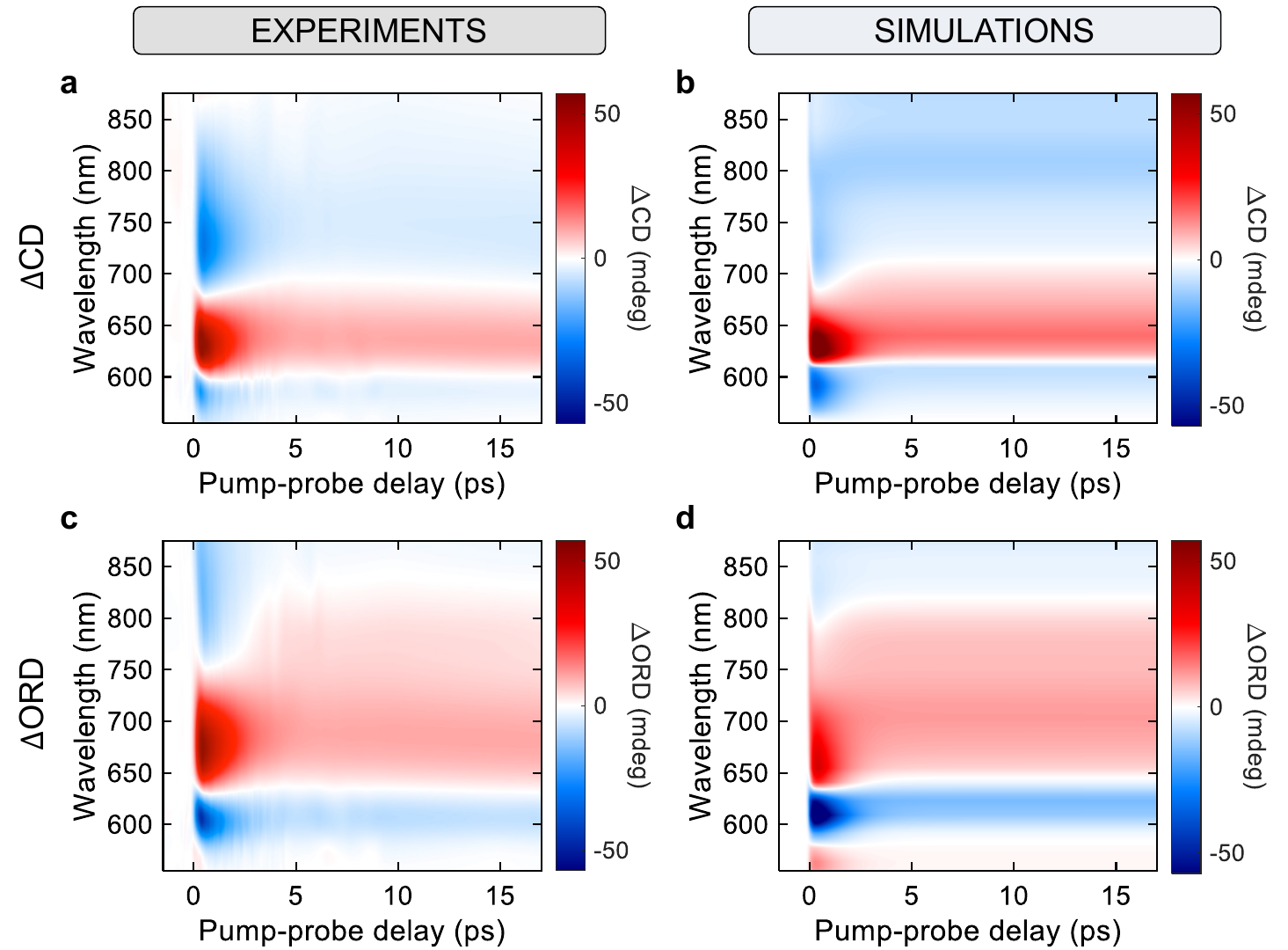}
\caption{\textbf{Ultrafast chiro-optical response of plasmonic helicoids. a, b.} Measured (\textbf{a}) and simulated (\textbf{b}) maps of the ultrafast $\Delta$CD, as a function of the probe wavelength and pump-probe delay. \textbf{c, d.} Measured (\textbf{c}) and simulated (\textbf{d}) maps of the ultrafast $\Delta$ORD, as a function of the probe wavelength and pump-probe delay. 
}\label{fig3}
\end{figure}

In the ultrafast, nonequilibrium regime, the absorption of the pump pulse induces nonlinear changes in the optical properties of the plasmonic nanostructures, mediated by the excitation of high-energy 'hot' carriers. 
These variations, in turn, lead to transient modulations of the system's chiral response, which we track experimentally and describe through our simulations. Figure~\ref{fig3} summarises the results of the broadband TR chiro-optical spectroscopy experiments. 
We introduce a linearly polarized pump pulse at \SI{515}{nm}  to photo-excite the system and record the pump-induced change in the chiral interferogram, thus obtaining delay-dependent interferometric maps as a function of the interferometer delay $t$ and pump-probe delay $\tau$. By applying a FT-based procedure similar to that used for the static measurements (see Supplementary Note 1), we obtain the TR-CD and TR-ORD spectra as a function of $\tau$ with sub-200-fs temporal resolution, limited by the pump pulse duration. 

\begin{figure}[H]
\centering
\includegraphics[width=0.9537\textwidth]{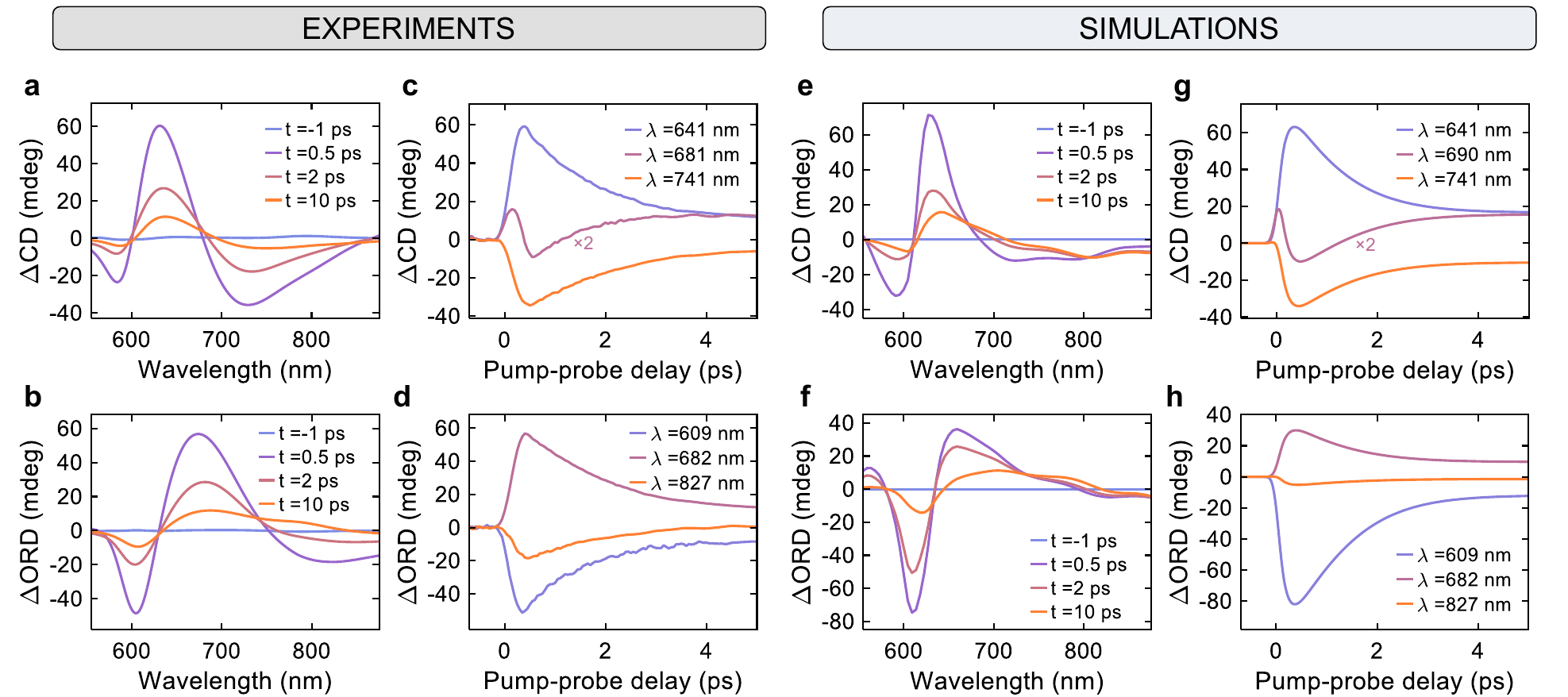}
\caption{\textbf{Spectro-temporal dynamics of chiral plasmonic helicoids. a, b.} Spectral cuts of the measured $\Delta$CD (\textbf{a}) and $\Delta$ORD (\textbf{b}) spectra at selected pump-probe delays $\tau$ = $\SI{-1}{ps}$ (blue), \SI{0.5}{ps} (purple), \SI{2}{ps} (red), \SI{10}{ps} (orange). \textbf{c, d.} Time traces of the measured $\Delta$CD (\textbf{c}) and $\Delta$ORD (\textbf{d}) at selected probe wavelengths $\lambda = \SI{641}{nm}$ (blue), \SI{681}{nm} (purple), \SI{741}{nm} (orange) for $\Delta$CD (\textbf{c}), $\lambda = \SI{609}{nm}$ (blue), \SI{682}{nm} (purple), \SI{827}{nm} (orange) for $\Delta$ORD (\textbf{d}). \textbf{e, f.} Same as \textbf{a, b} for the simulated $\Delta$CD (\textbf{e}) and $\Delta$ORD spectra (\textbf{f}). \textbf{g, h.} Same as (\textbf{c, d}) for the simulated $\Delta$CD (\textbf{g}, where a minor spectral shift is considered for the wavelength corresponding to the purple trace) and $\Delta$ORD (\textbf{h}) time traces.}\label{fig4}
\end{figure}

Figure \ref{fig3}a shows the map of $\Delta$CD($\lambda$,$\tau$), defined as $\Delta$CD = CD$_\text{ON}$ - CD$_\text{OFF}$, where CD$_\text{ON}$ (CD$_\text{OFF}$) refers to the pump-excited (unperturbed) CD signal, as a function of probe wavelength $\lambda$ and pump-probe delay $\tau$. 
After interaction with the pump pulse, $\Delta$CD shows a prominent positive peak between \SI{600}{nm} and \SI{680}{nm}, along with negative lobes at shorter and longer wavelengths.
For better visualisation, Fig.~\ref{fig4}a reports vertical sections of the $\Delta$CD map at selected delays, clearly showing the spectral structure of the transient signal, and its gradual decrease in amplitude as the pump-probe delay increases, from \SI{500}{fs} (purple) to \SI{2}{ps} (red) and \SI{10}{ps} (orange).
Over time, these lobes exhibit a peculiar evolution, with a pronounced redshift within the first $\sim \SI{200}{fs}$ after pump arrival and subsequent onset of the nonlinear modulations. 
This evolution is manifest when tracking the isosbestic lines (corresponding to the white shade in Fig.~\ref{fig3}a) in the maps over time, as well as inspecting temporal traces of $\Delta$CD at fixed wavelengths, shown in Fig.~\ref{fig4}c.
Specifically, while the dynamics extracted at the peaks of the $\Delta$CD (corresponding to \SI{641}{nm}, blue trace, and \SI{741}{nm}, orange trace) reach a maximum (in absolute value) around $\sim \SI{350}{fs}$ and decay back to equilibrium following the system's relaxation, the trace at \SI{681}{nm} (purple in Fig.~\ref{fig4}c) exhibits a double sign inversion, resulting from the non-trivial spectral modulations of the zero-crossing points in the $\Delta$CD map.
Similar arguments hold for the spectro-temporal evolution of $\Delta$ORD, whose full map is reported in Fig.~\ref{fig3}c, with its spectral and temporal cuts shown in Figs.~\ref{fig4}b and \ref{fig4}d, respectively.

Ultrafast experiments are supported by time-resolved simulations, which extend the modelling approach used in static conditions to the out-of-equilibrium regime.
We perform the same full-wave calculations and apply the same Jones calculus as in the unperturbed case, now incorporating the effects of photo-excitation in the plasmonic helicoids.
Upon pump absorption, hot carriers, i.e.~non-equilibrium electron-hole pairs with excess energy as high as the incident photons, are generated, persisting for a few picoseconds before equilibrating with the lattice via electron-phonon scattering. 
In the optical domain, the excited energy occupancy distribution of these carriers induces nonlinear variations of the allowed transitions in the metal, resulting in time- and probe-wavelength-dependent changes in the metal permittivity, $\Delta \varepsilon(\lambda, \tau)$. 
To describe these hot-carrier-mediated optical nonlinearities, we adopt a modelling approach previously reported \cite{schirato_transient_2020}, combining an inhomogeneous version of the Three-Temperature Model, which tracks the spatio-temporal dynamics of hot carriers, with a semi-classical treatment of Au thermo-modulational nonlinearities, to quantify $\Delta \varepsilon$ (see Methods Section 3 for details).
In terms of chiro-optical response, a given pump-induced $\Delta \varepsilon$ alters the absorption of LCP and RCP light to different extents.
This differential modulation results in transient variations of the system's inherent CD and ORD, captured as $\Delta$CD and $\Delta$ORD.

Simulation results for $\Delta$CD and $\Delta$ORD are shown in Figs.~\ref{fig3}b and \ref{fig3}d, respectively.
The calculated maps closely match the experiments, reproducing all key spectral and temporal features with accuracy.
This agreement is further highlighted in Fig.~\ref{fig4}, which compares the experiments with the spectral (Figs.~\ref{fig4}e and \ref{fig4}f versus Figs.~\ref{fig4}a and \ref{fig4}b, respectively) and temporal (Figs.~\ref{fig4}g and \ref{fig4}h versus Figs.~\ref{fig4}c and \ref{fig4}d, respectively) sections extracted from the simulated TR-CD and TR-ORD maps.
Notably, the simulation captures the peculiar dynamics of the isosbestic curves in the $\Delta$CD and $\Delta$ORD maps, and accurately reproduces the distinctive ultrafast evolution of the TR-CD at \SI{681}{nm} (compare purple traces in Figs.~\ref{fig4}g and \ref{fig4}c), also providing a physical explanation for this behaviour.
The model allows us to disentangle the contributions to the $\Delta$CD arising from lattice heating and from two distinct hot-electron populations: (i) `non-thermal' carriers, with a non-Fermi-Dirac energy distribution that relaxes within hundreds of femtoseconds, and (ii) `thermalised' carriers, described by a Fermi-Dirac distribution at an increased electronic temperature (see Methods Section 4).
In general, these contributions drive distinct spectro-temporal optical nonlinearities \cite{dellavalle_realtime_2012}.
For the plasmonic helicoids under study, their interplay underlies the observed double sign inversion in the transient CD (purple curve in Fig.~\ref{fig4}c), as confirmed by the maps shown in Supplementary Note 6.
As in the static case, minor discrepancies between simulations and experiments, more evident at longer wavelengths, are attributed to orientation disorder among helicoids in the sample. 
Nonetheless, the model successfully captures all dynamical features with accuracy and provides robust support to the experimental findings. 

To conclude our numerical analysis, we further investigate the TR-CD by inspecting the dynamic evolution of the electric and magnetic dipole moments supported by the helicoids. 
In particular, we track  their mutual interaction over time, which has been reported to contribute to CD through a term proportional to the imaginary part of their in-plane scalar product \cite{zhu_giant_2017} (see Supplementary Note 6 for a more detailed discussion).  
Notably, the transient differential variation of this quantity exhibits spectral features closely matching those of the $\Delta$CD, including the dominant positive peak centered at 650 nm, and the two weaker negative lobes at shorter and longer wavelengths.
The temporal dynamics of $\Delta$CD can likewise be qualitatively related to that of the dipolar interaction, both showing comparable rise and decay times, governed by the pump-induced permittivity change $\Delta\varepsilon$. 
These findings indicate that the main features of the transient variation of the helicoids' CD can be traced back to the ultrafast modulations of the interaction between dipole moments, driven by the hot-carrier-mediated optical nonlinearities, thereby linking near- and far-field quantities in the out-of-equilibrium regime.

\subsection{Spin-selective excitation in perovskites}
Our results thus far prove the ability to measure static and transient CD and ORD spectra in chiral samples. 
To further explore the capabilities of our experimental approach, we move to an achiral sample, where a chiro-optical response can be induced by illumination with circularly polarized light.
To this end, we selected a hybrid lead mixed-halide perovskite (methylammonium (MA) lead iodide/bromide: MAPb(I$_{0.7}$Br$_{0.3}$)$_3$), where the presence of heavy atoms significantly enhances spin-orbit coupling (SOC). 

\begin{figure}[H]
\centering
\includegraphics[width=0.8992\textwidth]{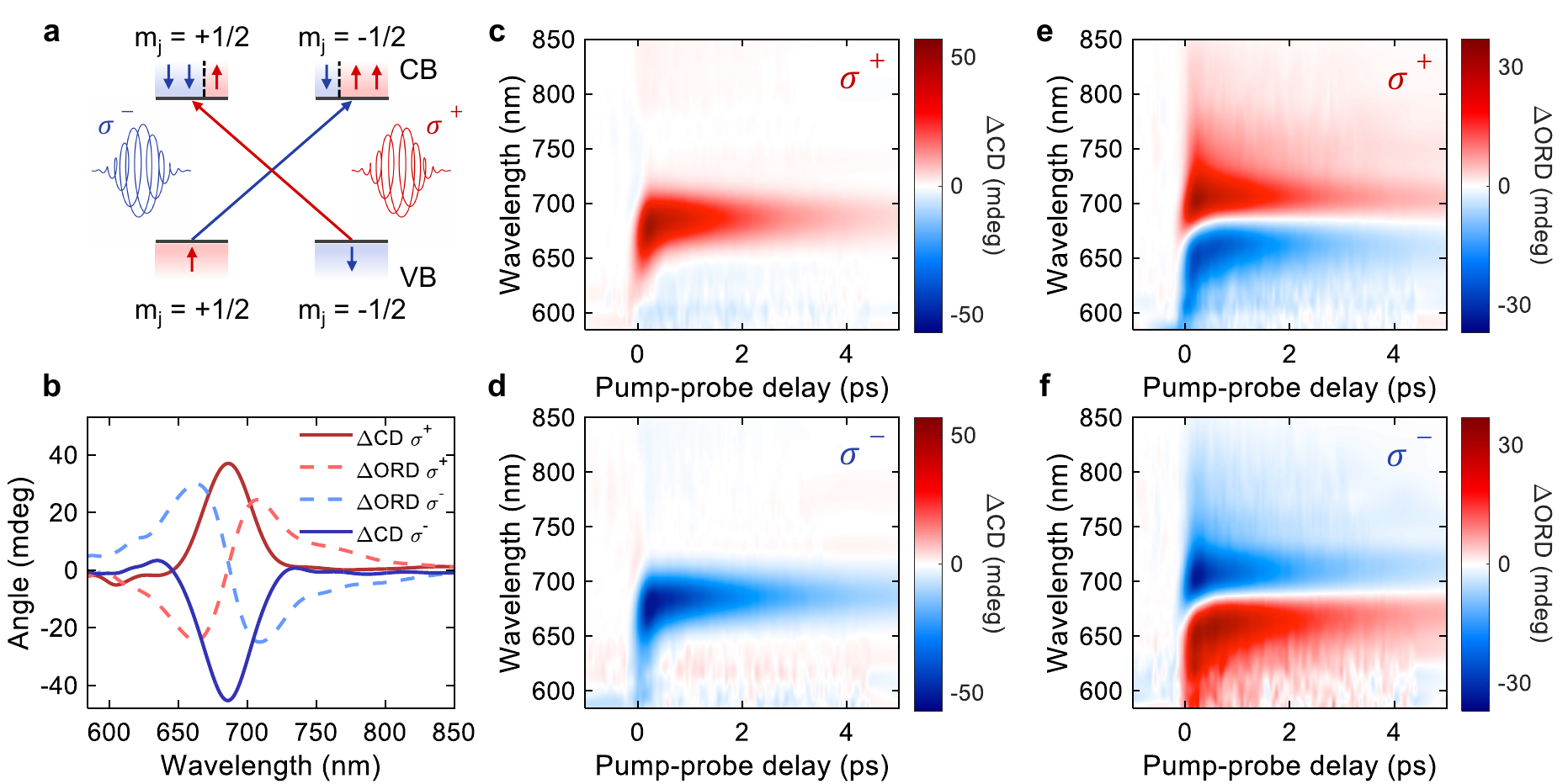}
\caption{\textbf{Ultrafast chiro-optical spectroscopy on lead halide perovskites . a.} Scheme of the electronic bands close to the band edge, and the spin-polarized populations of electrons and holes upon excitation with circularly polarized light. \textbf{b.} TR-CD (solid lines) and TR-ORD (dashed lines) spectra recorded at 1 ps pump-probe delay for different pump light helicities (solid and dashed red lines for $\sigma^+$, solid and dashed blue lines for $\sigma^-$). \textbf{c.} TR-CD map upon $\sigma^+$ photoexcitation. \textbf{d.} Same as (\textbf{c}) for opposite pump pulse helicity. \textbf{e.} TR-ORD map upon $\sigma^+$ photoexcitation. \textbf{f.} Same as (\textbf{e}) for opposite pump pulse helicity. The employed pump fluence was 15 $\mu$J/$\mathrm{cm}^2$}\label{fig5}
\end{figure}

This strong SOC causes the conduction band (CB) to split into two states with different total angular momentum $J$: a lower state with $J$=1/2 and an upper state with $J$=3/2, while the valence band (VB) remains largely unaffected. At the band edge, the VB has an orbital angular momentum $L$=0 (leading to m$_j$ = m$_s$), whereas the CB is characterized by $L$=1 and $J$=1/2. According to optical selection rules, circularly polarized photo-excitation conserves $J$, with transitions obeying $\Delta$m$_j$ = $\pm$ 1. Absorption of a photon with $\sigma^+$ ($\sigma^-$) helicity results in $\Delta$m$_j$ = $+$ 1 ($\Delta$m$_j$ = $-$ 1), promoting spin-down (spin-up) electrons which acquire a final m$_j$ =$+1/2$ (m$_j$ =$-1/2$) and leave behind spin-down (spin-up) holes, as shown in Fig. \ref{fig5}a. In the m$_j$ =$+1/2$ states, the spin-down population exceeds the spin-up population by a 2:1 ratio, with the reverse occurring for the m$_j$ =$-1/2$ states\cite{giovanni_highly_2015}. 

Consequently, by illuminating the sample with circularly polarized light, we expect to generate spin-polarized charge carriers, leading to a different absorption of the left and right circular components of the probe beam and to $\Delta$CD and $\Delta$ORD spectra with opposite signs for opposite handedness of the pump pulse. Figure \ref{fig5}b shows $\Delta$CD and $\Delta$ORD spectra for $\sigma^+$ and $\sigma^-$ excitation, measured at 1 ps pump-probe delay. For RCP pump, the $\Delta$CD spectrum shows a positive peak centered at the optical bandgap of the perovskite, while the $\Delta$ORD has a derivative shape, as expected according to the Kramers-Kronig relationship. Switching the helicity of the pump to LCP, we obtain the same spectral shape for both $\Delta$CD and $\Delta$ORD, this time with a perfect flip in sign with respect to zero.

 Figures~\ref{fig5}c and \ref{fig5}d report the $\Delta$CD maps for opposite pump helicities. Immediately after excitation, the signal exhibits a broader peak, distributed towards shorter wavelengths, then it rapidly redshifts within a few hundred femtoseconds. We attribute this behavior to the thermalization and cooling of spin-polarized hot carriers\cite{dai_thermalization_2023}. Circularly polarized light, tuned above resonance, generates a non-Fermi-Dirac spin-polarized population at energies higher than the optical bandgap. This non-equilibrium population rapidly thermalizes, on a timescale faster then our temporal resolution. 
 Afterwards, the hot carriers relax toward the band edge through scattering events that alter both energy and momentum (due to the parabolic nature of the electronic bands), resulting in the observed redshift of the CD peak.

During thermalization and cooling, spin polarization is partially lost, predominantly through the Elliott-Yafet mechanism, where the spin relaxation rate scales with the momentum scattering rate and the spin polarization decreases exponentially with the excess energy with respect to the band edge\cite{giovanni_ultrafast_2019}. Despite the large 0.6-eV excess energy of our pump pulses, we still observe a sizeable signal of $\approx$50 mdeg, that fully recovers in $\approx$10 ps. As elucidated by Giovanni et al.\cite{giovanni_highly_2015}, the relaxation process is expected to be biexponential, where the fast ($\tau_h$) and slow ($\tau_e$) decay constants are attributed to hole and electrons $J$-flip rates respectively. The hole $J$-polarization loss was found to be approximately one order of magnitude faster than the electron one and is below our temporal resolution ($\tau_h \lesssim$ 200 fs). On the other hand, we clearly resolve the electron spin polarization process, with a  $\tau_e$ = 2 ps.
Both these intraband angular momentum flips are found to be much faster than electron-hole recombination, which in this material proceeds on the nanosecond timescale. Figures \ref{fig5}e and \ref{fig5}f show the corresponding $\Delta$ORD maps. By tracing the isosbestic point and fitting it with a mono-exponential function, we can precisely infer the cooling time constant, finding $\tau_\text{c}$ = 300 fs (see Supplementary Note 7).

\section*{Discussion and Conclusions}\label{sec3}

We have presented an innovative broadband ultrafast chiro-optical spectroscopy technique, combining sensitive measurement of the photoinduced polarization rotation of a probe pulse via a balanced detector with Fourier transform spectroscopy to achieve frequency resolution. The setup has a remarkably simple architecture which employs a common-path  birefringent interferometer to delay, with high precision, chiral and achiral free-induction decay fields and, thanks to the self-heterodyned interferometric approach, detects both real and imaginary parts of the transient chiro-optical susceptibility. 

We demonstrate the detection of millidegree-level $\Delta$CD/$\Delta$ORD signals, acquired in a few tens of seconds, and we push the sensitivity down to tens of microdegrees by increasing the integration time. Our current spectral coverage  (550-950 nm) spans the visible and near-infrared ranges, but can be easily extended to the mid infrared and UV regions by suitable modifications of the source and detection optics. 
Moreover, the flexibility of our detection scheme allows for more advanced experiments, e.g. by addition of novel pump schemes, having non-trivial polarization states or pulse sequences for coherent control. Another viable option is its implementation in two-dimensional (2D) electronic/vibrational spectroscopy setups, allowing  2D chiro-optical spectroscopy (2D-CD/2D-ORD)\cite{fidler_dynamic_2014} and opening unprecedented possibilities in the detection of enantioselective interactions in molecules, and chiral or valley selective couplings in quantum materials.

\section*{Methods}\label{sec11}
\subsection{Gold nano-helicoids array fabrication}
The fabrication of the sample involves three subsequent steps, as described elsewhere \cite{lee_amino-acid-_2018,im_synthesis_2025}.
First, octahedral Au nanoparticles are fabricated via colloidal synthesis, and serve as seed for the nucleation of the helicoids. The seed nanoparticles are centrifuged at 12,000 rpm for 3 minutes and then dispersed in a 1 mM CTAB aqueous solution. 
The growth of the chiral nanoparticles is achieved through colloidal synthesis with 100 µL of gold chloride trihydrate as metal precursor, and 800 µL of 100 mM CTAB as chiral modifier, dissolved in  3.95 mL of deionized water. The reduction of Au$^{3+}$ was initiated by quickly injecting 475 µL of a 100 mM ascorbic acid solution. The nanoparticle growth was then triggered by adding 5 µL of l-GSH (GSH) at various concentrations, followed by the addition of 50 µL of the octahedral seed solution. The growth solution was incubated for 2 hours in a 30$^{\circ}$C water bath. Afterward, the solution was centrifuged three times at 6000 rpm for 1 minute and then redispersed in a 1 mM CTAB aqueous solution.

Lastly, the ordered array of helicoids is prepared through periodical assembling in a PDMS substrate. The substrate is prepared so as to contain an ordered array of nanowells with 400 nm pitch parameters using nano-imprinting methods\cite{kim_enantioselective_2022}. Deposition is performed via dual phase assembly. A thin monolayer of hexane containing dispersed chiral nanoparticles is distributed on deionized water. The PDMS substrate is then immersed in the monolayer and dip coated with the helicoid dispersion. The helicoids are finally allocated in the nanowells by teflon stick rolling.

\subsection{Perovskite thin film fabrication}

The perovskite thin film was fabricated by spin-coating a precursor solution. This precursor was prepared from stock solutions of MAPbI$_3$ (0.7 M, MAI:PbI$_2$ = 1:1) and MAPbBr$3$ (0.7 M, MABr:PbBr$2$ = 1:1), using dimethyl sulfoxide (DMSO) and dimethylformamide (DMF) as solvents in a 9.29:0.71 molar ratio. The two stocks were then mixed in the appropriate stoichiometric ratio to yield the MAPb(I${0.7}$Br${0.3}$)$_3$ formulation.
The mixed formulation was stirred at 80 $^\circ$C for 30 min, followed by stirring overnight (12 h) at room temperature inside the glovebox. 

The perovskite precursor solution (MAPb(I${0.7}$Br${0.3}$)$_3$) was deposited by spin coating at 4000 rpm for 25 s. During the spin coating, 200 µL of ethyl acetate (antisolvent) was dripped 8 seconds after the start of process. The film was then annealed at 100 $^\circ$C for 10 min inside the glovebox. After cooling, the perovskite was encapsulated with polymethyl methacrylate (PMMA) by spinning it at 3000 rpm for 60 s, followed by annealing at 80 $^\circ$C for 2 min. The encapsulating solution was previously obtained by adding 50 mg/ml of PMMA in chlorobenzene, stirring at room temperature until completely dissolved.

\subsection{Transient chiro-optical spectroscopy}
The experimental setup starts with a regeneratively amplified Yb:KGW laser (Pharos, Light Conversion), delivering 200 fs pulses at 1030 nm wavelength with 200 µJ energy and 100 kHz repetition rate. A fraction of the laser output is split into two replicas  to generate the pump and probe pulses. The pump beam is modulated at 50 kHz by means of an electro-optic modulator and a polarizing beam splitter, and the 515 nm pump pulses are generated by frequency doubling in a $\beta$-barium borate crystal and filtering through a 650 nm shortpass filter. A motorized delay line on the pump beam is used to control the pump-probe delay. The broadband probe pulse is a white light continuum (WLC) obtained by focusing the 1030-nm beam on a 10-mm-thick YAG plate and filtered via a 950 nm shortpass filter, thus obtaining a WLC extending from 550 to 950 nm. A Glan-Taylor (GT) polarizer with 1:100000 extinction rate is then added to the probe optical path so as to ensure a clean, linear polarization state. 

Both pump and probe beams are focused non collinearly on the sample. The pump spot size ($1/e$ diameter) is approximately  170 µm, and is approximately twice as large as the probe spot size, ensuring a uniform illumination of the probed region The probe is then collected and sent to the CPI, which is a modified version of the Gemini interferometer (NIREOS srl), where both polarizers were removed. In order to avoid spurious achiral signals, the GT polarizer is precisely rotated so as to align its polarization axis with the extraordinary axis of the wedges in the CPI. The GT is set so as to give a flat interferogram in absence of the sample, i.e. in absence of any orthogonally polarized CFID. We are hence sure that no delayed pulse replicas are generated by the CPI, which would give a spurious signal. The output of the CPI is coupled to a polarization bridge, consisting in a Wollaston prism (WP) and a Si-based balanced photodiode (BPD) followed by a differential amplifier. 

The detection scheme is based on a LIA. The output of the BPD is demodulated at 50 kHz (i.e. the repetition rate of the pump beam, and half the repetition rate of the probe beam), so that the measured output directly provides a differential signal. The phase of the demodulation signal is optimized by sending the attenuated pump beam on a photodiode and maximinzing the LIA reading. For every pump-probe delay, we record full differential interferograms by scanning the wedge position of the CPI. 

The achiral interferogram, which is necessary to quantitatively calculate the light-induced chiral susceptibility, is obtained by acquiring a static interferogram (demodulating the signal at 100 kHz) when the polarizer is set to 45° with respect to the optical axis of the CPI. A normalization factor is introduced to compensate for the higher losses due to the different polarizer angle. In this case, since the intensities of the ordinary and extraordinary beams are equal, the CPI acquires the phase and amplitude of the probe spectrum.

The full interferometric map is then apodized, Fourier transformed, and divided by the achiral calibration spectrum (see also Supplementary Note 1). The real (imaginary) part of the FT of the interferogram corresponds to the delay dependent $\Delta$ORD ($\Delta$CD) spectrum.

\subsection{Numerical modelling of chiral Au nanoparticles}

The sample’s chiro-otical response was modelled by retrieving the expressions for ORD and CD via Jones calculus. In the small dichroism approximation (see Supplementary Note 5 for further details), we obtained:

\begin{gather}
    \text{ORD} = -\operatorname{Im}\Big({\frac{t_{LL}-t_{RR}}{t_{LL}+t_{RR}}}\Big),\\
    \text{CD} = -\operatorname{Re}\Big({\frac{t_{LL}-t_{RR}}{t_{LL}+t_{RR}}}\Big).
\end{gather}

\noindent
Here $t_{LL}$ and $t_{RR}$ are the diagonal elements of the Jones matrix written in circular basis, describing left and right co-polarized field transmission coefficients, respectively.

In static, unperturbed conditions, these complex-valued transmission coefficients were computed via full-wave electromagnetic simulations, using a finite-element-method (FEM)-based commercial software (COMSOL Multiphysics 6.2). The experimental sample was modeled as a perfect, infinite array of identical helicoids. Floquet periodic boundary conditions were imposed on the sides of the array unit cell, while the top/bottom edges (air and PDMS substrate, respectively) were set as input/output ports with a specific, either left (anti-clockwise, when observed from the source point of view, for a plane wave propagating along $-z$) or right (clockwise), polarization.
 
The geometrical configuration of the unit cell was adjusted to best fit the experimental static response (both CD and ORD). Specifically, we set an in-plane periodicity $p=\SI{405}{nm}$, close to the nominal value ($\SI{400}{nm}$), and a helicoid side length $L=\SI{212}{nm}$, slightly increased compared to the nominal one ($\SI{180}{nm}$). The nanostructure was partially embedded by $\SI{160}{nm}$ into the PDMS substrate (refractive index $n=1.5$), and rotated by $\ang{24}$ about the $x$-axis, by $\ang{16.5}$ about the $y$-axis. 
 The Au static permittivity within the helicoid domain was modeled with analytical Drude-Lorentz formulas \cite{etchegoin_analytic_2006} fitted on experimental data \cite{johnson_optical_1972}, with an increased Drude damping factor $\Gamma$ to account for the inhomogeneous broadening of the spectral features of the sample response introduced by size and shape dispersion \cite{husnik_quantitative_2012}. A four-fold increase of the bulk value \cite{etchegoin_analytic_2006} was considered to achieve good agreement with the experiments. 

The ultrafast transient chiro-optical response of the helicoid array was calculated by applying a multi-step semiclassical modeling approach. First, the photoinduced generation of hot carriers and their subsequent ultrafast dynamics were modeled via the well-established Three-Temperature Model (3TM), in its inhomogeneous formulation (I3TM) \cite{schirato_transient_2020}
The I3TM consists of a system of coupled rate equations that describe the ultrafast spatio-temporal carrier dynamics in plasmonic nanostructures under ultrashort laser pulse illumination in terms of three space-time dependent energetic degrees of freedom: (i) the energy density stored in an out-of-equilibrium population of non-thermal electrons $N$, (ii) the temperature of thermalized hot electrons $\Theta_E$ , and (iii) the metal nanostructure lattice temperature $\Theta_L$. These three spatio-temporal variables are coupled by the following equations:

\begin{gather}
    \frac{\partial N}{\partial t} = -aN-bN+P_\text{abs}\\
    C_E\frac{\partial \Theta_E}{\partial t}=-\nabla\cdot(-\kappa_E\nabla\Theta_E)-G(\Theta_E-\Theta_L)+aN\\
    C_L\frac{\partial \Theta_L}{\partial t}=\kappa_L\nabla^2\Theta_L+G(\Theta_E-\Theta_L)+bN
\end{gather}

\noindent
where the coefficients govern the energy relaxation following photoexcitation. Specifically, $a$ and $b$ regulate the relaxation of non-thermal electrons through electron-electron and electron-phonon scattering, respectively. $C_E$ and $C_L$ represent the heat capacities of thermalized electrons and lattice, while $\kappa_E$ and $\kappa_L$ are their respective thermal conductivities. Finally, $G$ is the electron-phonon coupling term. Further details on the inhomogeneous 3TM and its coefficients can be found elsewhere \cite{schirato_transient_2020}. Lastly, the driving term $P_\text{abs}$ represents the pump pulse electromagnetic power density absorbed by the nanostructure. It is expressed as:

\begin{equation}
P_\text{abs}(r,t) = A(\lambda_P,r)F_P\frac{S}{V}g(t)
\end{equation}

\noindent
 where $F_P=\SI{300}{\micro\joule/\cm\squared}$ is the pump fluence, $A(\lambda_P,r)$ is the pump absorption spatial pattern at the pump central wavelength $\lambda_P = \SI{515}{nm}$, estimated by FEM numerical simulations considering a monochromatic plane wave linearly polarized along the $x$-axis exciting the helicoid array.
 $S$ and $V$ are respectively the unit cell surface and nanoparticle volume, while $g(t)$ is a Gaussian profile with FWHM duration of $\SI{180}{fs}$, in agreement with the measurement conditions.

 The solution of the I3TM, retrieved via a segregated algorithm implemented in COMSOL, was combined with a semiclassical description of the optical nonlinearities in Au mediated by non-equilibrium hot carriers to compute the pump-induced transient variations of the metal permittivity. 
In brief, both $N$ and $\Theta_E$ contribute to modifications in the electron energy occupancy distribution, which in turn affect the metal optical absorption associated with interband transitions \cite{rosei_dbands_1973}. This leads to a nonlinear change in the imaginary part of the permittivity, with the corresponding real part determined by Kramers-Kronig relations. 
An increase in the lattice temperature is instead related to modifications in the Drude-like intraband term of the metal permittivity, due to changes in the plasma frequency \cite{yeshchenko_temperature_2013} and damping factor \cite{smith_frequency_1982}.
To compute the spectro-temporal permittivity modulation, the values of $N$, $\Theta_E$ and $\Theta_L$  were averaged over the volume of a single nanoparticle, considering these values as representative of the out-of-equilibrium state of the plasmonic nanostructure at each time instant.
Finally, starting from the time-dependent transient permittivity, we adopted a perturbative approach to compute the $\Delta$CD and $\Delta$ORD signals as a linear combination of the complex-valued permittivity modulation, weighted by spectral coefficients that we retrieved directly for the two chiro-optical quantities.



\subsection{Acknowledgements}

We acknowledge Franco V. A. Camargo for the valuable  discussions.

\subsection{Funding}

G.Ce., A.I. acknowledge support from the European Union’s NextGenerationEU Programme with the I-PHOQS Infrastructure [IR0000016, ID D2B8D520, CUP B53C22001750006] ``Integrated infrastructure initiative in Photonic and Quantum Sciences". 
A.S., G.D.V. acknowledge the European Union’s Horizon Europe research and innovation programme under the Marie Sk\l{}odowska-Curie Action PATHWAYS HORIZON-MSCA-2023-PF-GF grant agreement no.~101153856, and the METAFAST project that received funding from the European Union's Horizon 2020 Research and Innovation program under Grant Agreement No. 899673.

A.V.Y T., G. Cr., M.M.~acknowledge financial support from the ERC-StG ULYSSES grant agreement no.~101077181 funded by the European Union. Views and opinions expressed are however those of the author(s) only and do not necessarily reflect those of the European Union or the European Research Council. Neither the European Union nor the granting authority can be held responsible for them.

K.T.N. appreciated the support by the Nano \& Material Technology Development Program through the National Research Foundation of Korea (RS-2024-00409405). 

A.R. acknowledges the financial support from the project “nuovi Concetti, mAteriali e tecnologie per l'iNtegrazione del fotoVoltAico negli edifici in uno scenario di generazione diffuSa” [CANVAS], funded by the Italian Ministry of the Environment and the Energy Security, through the Research Fund for the Italian Electrical System (type-A call, published on G.U.R.I. n. 192 on 18-08-2022) CUP-B53C22005670005. 

D.A. acknowledges the support of the Italian Ministry of University and Research (MUR) through M.D. 351/2022 ("PNRR") Scholarship No. CUPF83C22000810006.

G.D.V. acknowledges the support from the HOTMETA project under the PRIN 2022 MUR program funded by the European Union – Next Generation EU - “PNRR - M4C2, investimento 1.1 - “Fondo PRIN 2022” - HOT-carrier METasurfaces for Advanced photonics (HOTMETA), contract no. 2022LENW33 - CUP: D53D2300229 0006”. 

\subsection{Conflict of interest}

G. Ce. declares association with NIREOS s.r.l..

\subsection{Data availability}

All data generated or analysed during this study are available from the corresponding author upon reasonable request.

\subsection{Author contribution}

G. Ce. conceived and supervised the project. F.G. and A.I. performed the experiments and analyzed the data with the support of A.V.. A.V.Y T. performed the simulations, with the contribution of A.S., under the supervision of G.D.V.. G. Cr. performed the analytical derivation of CD and ORD, under the supervision of M.M. and G.D.V.. R.M.K., S.M.L. and J-H.H. fabricated the gold nano-helicoids sample, under the supervision of K.T.N.. D.A. fabricated the perovskite sample, under the supervision of A.R.. All authors contributed to the discussion of the results and the writing of the paper.


\begin{thebibliography}{99}

\bibitem{ayuso_ultrafast_2022}
Ayuso, D., Ordonez, A. F. \& Smirnova, O. Ultrafast chirality: the road to efficient chiral measurements. \textit{Phys. Chem. Chem. Phys.} \textbf{24}, 26962--26991 (2022).

\bibitem{changenet_recent_2023}
Changenet, P. \& Hache, F. Recent advances in the development of ultrafast electronic circular dichroism for probing the conformational dynamics of biomolecules in solution. \textit{EPJ ST} \textbf{232}, 2117--2129 (2023).

\bibitem{rong_interaction_2023}
Rong, R. \textit{et al.} The Interaction of 2D Materials With Circularly Polarized Light. \textit{Adv. Sci.} \textbf{10}, 2206191 (2023).

\bibitem{bao_light-induced_2022}
Bao, C., Tang, P., Sun, D. \& Zhou, S. Light-induced emergent phenomena in 2D materials and topological materials. \textit{Nat. Rev. Phys.} \textbf{4}, 33--48 (2022).

\bibitem{mak_control_2012}
Mak, K. F., He, K., Shan, J. \& Heinz, T. F. Control of valley polarization in monolayer MoS$_2$ by optical helicity. \textit{Nat. Nanotechnol.} \textbf{7}, 494--498 (2012).

\bibitem{vitale_valleytronics:_2018}
Vitale, S. A. \textit{et al.} Valleytronics: opportunities, challenges, and paths forward. \textit{Small} \textbf{14}, 1801483 (2018).

\bibitem{langer_lightwave_2018}
Langer, F. \textit{et al.} Lightwave valleytronics in a monolayer of tungsten diselenide. \textit{Nature} \textbf{557}, 76--80 (2018).

\bibitem{hubener_engineering_2021}
Hübener, H. \textit{et al.} Engineering quantum materials with chiral optical cavities. \textit{Nat. Mater.} \textbf{20}, 438--442 (2021).

\bibitem{tay_terahertz_2025}
Tay, F. \textit{et al.} Terahertz chiral photonic-crystal cavities for Dirac gap engineering in graphene. \textit{Nat. Commun.} \textbf{16}, 5270 (2025).

\bibitem{huang_giant_2023}
Huang, M. \textit{et al.} Giant nonlinear Hall effect in twisted bilayer WSe$_2$. \textit{Natl. Sci. Rev.} \textbf{10}, nwac232 (2023).

\bibitem{zhu_creating_2024}
Zhu, H. \& Yakobson, B. I. Creating chirality in the nearly two dimensions. \textit{Nat. Mater.} \textbf{23}, 316--322 (2024).

\bibitem{chen_ultrafast_2021}
Chen, Z., Dong, G. \& Qiu, J. Ultrafast Pump–Probe Spectroscopy—A Powerful Tool for Tracking Spin–Quantum Dynamics in Metal Halide Perovskites. \textit{Adv. Quantum Technol.} \textbf{4}, 2100052 (2021).

\bibitem{giovanni_highly_2015}
Giovanni, D. \textit{et al.} Highly spin-polarized carrier dynamics and ultralarge photoinduced magnetization in CH$_3$NH$_3$PbI$_3$ perovskite thin films. \textit{Nano Lett.} \textbf{15}, 1553--1558 (2015).

\bibitem{giovanni_ultrafast_2019}
Giovanni, D. \textit{et al.} Ultrafast long-range spin-funneling in solution-processed Ruddlesden–Popper halide perovskites. \textit{Nat. Commun.} \textbf{10}, 3456 (2019).

\bibitem{briscoe_photogalvanic_2025}
Briscoe, J. \& Shi, J. Photogalvanic effects in non-centrosymmetric halide perovskites. \textit{Nat. Rev. Phys.} \textbf{7}, 270--279 (2025).

\bibitem{ye_aboveroomtemperature_2014}
Ye, H.-Y., Zhang, Y., Fu, D.-W. \& Xiong, R.-G. An above-room-temperature ferroelectric organo–metal halide perovskite: (3-pyrrolinium)(CdCl$_3$). \textit{Angew. Chem.} \textbf{126}, 11424--11429 (2014).

\bibitem{zhai_giant_2017}
Zhai, Y. \textit{et al.} Giant Rashba splitting in 2D organic–inorganic halide perovskites measured by transient spectroscopies. \textit{Sci. Adv.} \textbf{3}, e1700704 (2017).

\bibitem{zheng_rashba_2015}
Zheng, F., Tan, L. Z., Liu, S. \& Rappe, A. M. Rashba spin–orbit coupling enhanced carrier lifetime in CH$_3$NH$_3$PbI$_3$. \textit{Nano Lett.} \textbf{15}, 7794--7800 (2015).

\bibitem{fan_chiral_2012}
Fan, Z. \& Govorov, A. O. Chiral nanocrystals: plasmonic spectra and circular dichroism. \textit{Nano Lett.} \textbf{12}, 3283--3289 (2012).

\bibitem{kwon_chiral_2023}
Kwon, J., Park, K. H., Choi, W. J., Kotov, N. A. \& Yeom, J. Chiral spectroscopy of nanostructures. \textit{Acc. Chem. Res.} \textbf{56}, 1359--1372 (2023).

\bibitem{luo_plasmonic_2017}
Luo, Y. \textit{et al.} Plasmonic chiral nanostructures: chiroptical effects and applications. \textit{Adv. Opt. Mater.} \textbf{5}, 1700040 (2017).

\bibitem{hentschel_chiral_2017}
Hentschel, M., Schäferling, M., Duan, X., Giessen, H. \& Liu, N. Chiral plasmonics. \textit{Sci. Adv.} \textbf{3}, e1602735 (2017).

\bibitem{kim_enantioselective_2022}
Kim, R. M. \textit{et al.} Enantioselective sensing by collective circular dichroism. \textit{Nature} \textbf{612}, 470--476 (2022).

\bibitem{link_virtual_2021}
Link, S. \& Hartland, G. V. Virtual issue on chiral plasmonics. \textit{J. Phys. Chem. C} \textbf{125}, 10175--10178 (2021).

\bibitem{wu_chiral_2022}
Wu, W. \& Pauly, M. Chiral plasmonic nanostructures: recent advances in their synthesis and applications. \textit{Mater. Adv.} \textbf{3}, 186--215 (2022).

\bibitem{lesko_optical_2024}
Lesko, D. M. B. \textit{et al.} Optical control of electrons in a Floquet topological insulator. Preprint at arXiv:2407.17917 (2024).

\bibitem{mitra_light-wave-controlled_2024}
Mitra, S. \textit{et al.} Light-wave-controlled Haldane model in monolayer hexagonal boron nitride. \textit{Nature} \textbf{628}, 752--757 (2024).

\bibitem{tyulnev_valleytronics_2024}
Tyulnev, I. \textit{et al.} Valleytronics in bulk MoS$_2$ with a topologic optical field. \textit{Nature} \textbf{628}, 746--751 (2024).

\bibitem{nova_effective_2017}
Nova, T. F. \textit{et al.} An effective magnetic field from optically driven phonons. \textit{Nat. Phys.} \textbf{13}, 132--136 (2017).

\bibitem{basini_terahertz_2024}
Basini, M. \textit{et al.} Terahertz electric-field-driven dynamical multiferroicity in SrTiO$_3$. \textit{Nature} \textbf{628}, 534--539 (2024).

\bibitem{rhee_coherent_2012}
Rhee, H., Eom, I., Ahn, S.-H. \& Cho, M. Coherent electric field characterization of molecular chirality in the time domain. \textit{Chem. Soc. Rev.} \textbf{41}, 4457 (2012).

\bibitem{jasperson_improved_1969}
Jasperson, S. N. \& Schnatterly, S. E. An improved method for high reflectivity ellipsometry based on a new polarization modulation technique. \textit{Rev. Sci. Instrum.} \textbf{40}, 761--767 (1969).

\bibitem{lewis_new_1985}
Lewis, J. W. \textit{et al.} New technique for measuring circular dichroism changes on a nanosecond time scale. Application to (carbonmonoxy)myoglobin and (carbonmonoxy)hemoglobin. \textit{J. Phys. Chem.} \textbf{89}, 289--294 (1985).

\bibitem{xie_picosecond_1989}
Xie, X. \& Simon, J. D. Picosecond time-resolved circular dichroism spectroscopy: experimental details and applications. \textit{Rev. Sci. Instrum.} \textbf{60}, 2614--2627 (1989).

\bibitem{hache_multiscale_2021}
Hache, F. \& Changenet, P. Multiscale conformational dynamics probed by time-resolved circular dichroism from seconds to picoseconds. \textit{Chirality} \textbf{33}, 747--757 (2021).

\bibitem{meyer-ilse_recent_2013}
Meyer-Ilse, J., Akimov, D. \& Dietzek, B. Recent advances in ultrafast time-resolved chirality measurements: perspective and outlook: Ultrafast transient molecular chirality. \textit{Laser Photonics Rev.} \textbf{7}, 495--505 (2013).

\bibitem{oppermann_broad-band_2019}
Oppermann, M. \textit{et al.} Broad-Band Ultraviolet CD Spectroscopy of Ultrafast Peptide Backbone Conformational Dynamics. \textit{J. Phys. Chem. Lett.} \textbf{10}, 2700--2705 (2019).

\bibitem{oppermann_chiral_2022}
Oppermann, M., Zinna, F., Lacour, J. \& Chergui, M. Chiral control of spin-crossover dynamics in Fe(II) complexes. \textit{Nat. Chem.} \textbf{14}, 739--745 (2022).

\bibitem{bourelle_optical_2022}
Bourelle, S. A. \textit{et al.} Optical control of exciton spin dynamics in layered metal halide perovskites via polaronic state formation. \textit{Nat. Commun.} \textbf{13}, 3320 (2022).

\bibitem{gucci_ultrafast_2024}
Gucci, F. \textit{et al.} Ultrafast valleytronic logic operations. Preprint at arXiv:2412.08318 (2024).

\bibitem{dal_conte_ultrafast_2015}
Dal Conte, S. \textit{et al.} Ultrafast valley relaxation dynamics in monolayer MoS$_2$ probed by nonequilibrium optical techniques. \textit{Phys. Rev. B} \textbf{92}, 235425 (2015).

\bibitem{crotti_giant_2024}
Crotti, G. \textit{et al.} Giant ultrafast dichroism and birefringence with active nonlocal metasurfaces. \textit{Light Sci. Appl.} \textbf{13}, 204 (2024).

\bibitem{zeng_photo-induced_2025}
Zeng, Z. \textit{et al.} Photo-induced chirality in a nonchiral crystal. \textit{Science} \textbf{387}, 431--436 (2025).

\bibitem{trifonov_broadband_2010}
Trifonov, A., Buchvarov, I., Lohr, A., Würthner, F. \& Fiebig, T. Broadband femtosecond circular dichroism spectrometer with white-light polarization control. \textit{Rev. Sci. Instrum.} \textbf{81}, 043104 (2010).

\bibitem{scholz_ultrafast_2019}
Scholz, M. \textit{et al.} Ultrafast Broadband Transient Absorption and Circular Dichroism Reveal Relaxation of a Chiral Copolymer. \textit{J. Phys. Chem. Lett.} \textbf{10}, 5160--5166 (2019).

\bibitem{oppermann_ultrafast_2019}
Oppermann, M. \textit{et al.} Ultrafast broadband circular dichroism in the deep ultraviolet. \textit{Optica} \textbf{6}, 56 (2019).

\bibitem{eom_single-shot_2012}
Eom, I., Ahn, S.-H., Rhee, H. \& Cho, M. Single-Shot Electronic Optical Activity Interferometry: Power and Phase Fluctuation-Free Measurement. \textit{Phys. Rev. Lett.} \textbf{108}, 103901 (2012).

\bibitem{niezborala_measuring_2006}
Niezborala, C. \& Hache, F. Measuring the dynamics of circular dichroism in a pump–probe experiment with a Babinet–Soleil compensator. \textit{J. Opt. Soc. Am. B} \textbf{23}, 2418 (2006).

\bibitem{rhee_femtosecond_2009}
Rhee, H., June, Y.-G., Lee, J.-S. \textit{et al.} Femtosecond characterization of vibrational optical activity of chiral molecules. \textit{Nature} \textbf{458}, 310--313 (2009).

\bibitem{eom_broadband_2011}
Eom, I., Ahn, S.-H., Rhee, H. \& Cho, M. Broadband near-UV to visible optical activity measurement using self-heterodyned method. \textit{Opt. Express} \textbf{19}, 10017 (2011).

\bibitem{hiramatsu_communication:_2015}
Hiramatsu, K. \& Nagata, T. Communication: Broadband and ultrasensitive femtosecond time-resolved circular dichroism spectroscopy. \textit{J. Chem. Phys.} \textbf{143}, 121102 (2015).

\bibitem{preda_time-domain_2018}
Preda, F., Perri, A., Réhault, J., Dutta, B., Helbing, J., Cerullo, G. \& Polli, D. 
Time-domain measurement of optical activity by an ultrastable common-path interferometer. 
\textit{Opt. Lett.} \textbf{43}, 1882 (2018).

\bibitem{ghosh_broadband_2021}
Ghosh, S., Herink, G., Perri, A., Preda, F., Manzoni, C., Polli, D. \& Cerullo, G. 
Broadband optical activity spectroscopy with interferometric Fourier-transform balanced detection. 
\textit{ACS Photonics} \textbf{8}, 2234–2242 (2021).

\bibitem{brida_phase-locked_2012}
Brida, D., Manzoni, C. \& Cerullo, G. Phase-locked pulses for two-dimensional spectroscopy by a birefringent delay line. \textit{Opt. Lett.} \textbf{37}, 3027 (2012).

\bibitem{lee_amino-acid-_2018}
Lee, H.-E. \textit{et al.} Amino-acid- and peptide-directed synthesis of chiral plasmonic gold nanoparticles. \textit{Nature} \textbf{556}, 360–365 (2018).

\bibitem{barron_molecular_2004}
Barron, L. D. \textit{Molecular Light Scattering and Optical Activity}, 2nd edn. (Cambridge Univ. Press, Cambridge, 2004).

\bibitem{ahn_highly_2024}
Ahn, H.-Y. \textit{et al.} Highly chiral light emission using plasmonic helicoid nanoparticles. \textit{Adv. Opt. Mater.} \textbf{12}, 2400699 (2024).

\bibitem{kim_controlling_2021}
Kim, J. W. \textit{et al.} Controlling the size and circular dichroism of chiral gold helicoids. \textit{Mater. Adv.} \textbf{2}, 6988–6995 (2021).

\bibitem{tang_enhanced_2011}
Tang, Y. \& Cohen, A. E. Enhanced enantioselectivity in excitation of chiral molecules by superchiral light. \textit{Science} \textbf{332}, 333–336 (2011).

\bibitem{schirato_transient_2020}
Schirato, A. \textit{et al.} Transient optical symmetry breaking for ultrafast broadband dichroism in plasmonic metasurfaces. \textit{Nat. Photonics} \textbf{14}, 723–727 (2020).

\bibitem{dellavalle_realtime_2012}
Della Valle, G., Conforti, M., Longhi, S., Cerullo, G. \& Brida, D. Real-time optical mapping of the dynamics of nonthermal electrons in thin gold films. \textit{Phys. Rev. B} \textbf{86}, 155139 (2012).

\bibitem{zhu_giant_2017}
Zhu, A. Y. \textit{et al.} Giant intrinsic chiro-optical activity in planar dielectric nanostructures. \textit{Light Sci. Appl.} \textbf{7}, 17158 (2017).

\bibitem{dai_thermalization_2023}
Dai, L., Ye, J. \& Greenham, N. C. Thermalization and relaxation mediated by phonon management in tin–lead perovskites. \textit{Light Sci. Appl.} \textbf{12}, 208 (2023).

\bibitem{fidler_dynamic_2014}
Fidler, A. F., Singh, V. P., Long, P. D., Dahlberg, P. D. \& Engel, G. S. Dynamic localization of electronic excitation in photosynthetic complexes revealed with chiral two-dimensional spectroscopy. \textit{Nat. Commun.} \textbf{5}, 3286 (2014).

\bibitem{im_synthesis_2025}
Im, S. W. \textit{et al.} Synthesis of chiral gold helicoid nanoparticles using glutathione. \textit{Nat. Protoc.} \textbf{20}, 1082–1096 (2025).

\bibitem{etchegoin_analytic_2006}
Etchegoin, P. G., Le Ru, E. C. \& Meyer, M. An analytic model for the optical properties of gold. \textit{J. Chem. Phys.} \textbf{125}, 164705 (2006).

\bibitem{johnson_optical_1972}
Johnson, P. B. \& Christy, R. W. Optical constants of the noble metals. \textit{Phys. Rev. B} \textbf{6}, 4370–4379 (1972).

\bibitem{husnik_quantitative_2012}
Husnik, M. \textit{et al.} Quantitative experimental determination of scattering and absorption cross-section spectra of individual optical metallic nanoantennas. \textit{Phys. Rev. Lett.} \textbf{109}, 233902 (2012).

\bibitem{rosei_dbands_1973}
Rosei, R., Antonangeli, F. \& Grassano, U. M. d bands position and width in gold from very low temperature thermomodulation measurements. \textit{Surf. Sci.} \textbf{37}, 689–699 (1973).

\bibitem{yeshchenko_temperature_2013}
Yeshchenko, O., Bondarchuk, I. S., Gurin, V., Dmitruk, I. \& Kotko, A. Temperature dependence of the surface plasmon resonance in gold nanoparticles. \textit{Surf. Sci.} \textbf{608}, 275–281 (2013).

\bibitem{smith_frequency_1982}
Smith, J. B. \& Ehrenreich, H. Frequency dependence of the optical relaxation time in metals. \textit{Phys. Rev. B} \textbf{25}, 923–930 (1982).


\end{thebibliography}

\section*{References}
\begingroup
\setlength{\leftskip}{1em}

\endgroup


\end{document}